\documentclass[12pt]{article}
\usepackage{amssymb,amsmath,amsfonts,multirow,color,graphicx}

\usepackage{tikz,lmodern}
\usepackage[most]{tcolorbox}

\usepackage{booktabs}
\usepackage{siunitx}
\usepackage{multirow}

\usepackage{acronym}
\acrodef{irn}[IRN]{Italian Railway Network}
\acrodef{crn}[CRN]{Chinese Railway Network}

\usepackage[bookmarks]{hyperref}
\usepackage{bookmark}

\title{Systemic collapse emergence from climate-induced perturbations in transportation networks}

\author
{Mauro Faccin$^{1,2*}$, Tomas Scagliarini $^{1,2}$, Manlio De Domenico $^{1,2\ast}$\\
\\
\normalsize
$^{1}$Depart of Physics and Astronomy ``Galileo Galilei'', Univ of Padova, Italy\\
\normalsize
$^{2}$Istituto Nationale di Fisica Nucleare (INFN), Sez di Padova, Padova, Italy\\
\normalsize
$^\ast$Corresponding authors. E-mail: mauro.faccin@unipd.it, manlio.dedomenico@unipd.it
}

\date{}

\begin{document}

\maketitle


\begin{abstract}
	The interaction between human infrastructures and external events results in loss of functionality or even complete system disruption through a snowball effect of cascading-failures.
	Their complex interplay challenges their prevention, affecting the human population that rely on their functionality.
	Understanding how these networks respond to external perturbations is crucial for predicting and mitigating the impacts of disruptive events.
	In particular, meteorological events, while not preventable, can have significant impacts on human infrastructure.
	Here, we propose a model coupling the dynamics of transportation systems to the dynamics of external events considered as fields, to capture at the same time the critical behavior of the former and the level of influence of the latter.
	The impact of rainfalls on real-world railway networks provides a compelling case study, showcasing local endemic and system-wide disruptions around a critical value of the rainfall strength.
	Remarkably, localized perturbations on the wider network reveals the local weaknesses of the global infrastructure, and we assess them in terms of system perturbability and corresponding societal risk. Our framework opens the door to the quantitative analysis of criticalities in essential human infrastructure imputable to vulnerabilities that are set to intensify due to climate change.
\end{abstract}

\section{Introduction}%

Human-made systems consist of a large number of interacting components whose nonlinear interactions give rise to emergent, often unpredictable, behaviors.
These systems can be conveniently modeled as complex networks~\cite{newman2018mathematics}, a mathematical framework that enables the analysis of interdependencies within these systems, ultimately improving our ability to manage and optimize them~\cite{artime2024robustness}. 
Examples span various domains, including transportation infrastructures, such as roads~\cite{Gallotti_2016}, air~\cite{Wu_2019} and rail networks~\cite{Tiong_2023_train_delay_review}, supply chains~\cite{Wiedmer_2021_supplychain} and financial markets~\cite{Battiston_2012_debtrank}.

The functionality of these networks can be disrupted by either structural shocks, such as the loss of nodes or links, or functional perturbations, affecting the internal state dynamics of those elements.
In power grids, for instance, local failures can propagate throughout the system, producing large-scale blackouts~\cite{Dobson_2007_blackouts, Chen_2020_nonlinear_cascades, Scagliarini_2025_fragility}.
This phenomenon, known as cascading failure~\cite{Motter_2002_cascades, zhao2016spatiotemporal}, is a paradigmatic example of systemic fragility in interconnected systems.
The network robustness analysis aims to quantify the ability of a system to withstand internal failures, external shocks, or external attacks~\cite{artime2024robustness, Motter_2004_cascades}, revealing phase transitions between functional and non-functional regimes~\cite{Dorogovtsev_2008_critical, Scheffer_2012_critical_transitions, Liu_2022_resilience}.
Attacks are often simulated in terms of random removal of nodes and links, or by informed attacks based on topological features~\cite{De_Domenico_2014_coverage}, with successful applications in medicine~\cite{Smart_2008_Cascading_metabolic, Bullmore_2009_brain}, ecology~\cite{Holling_1973_resilience_ecological_systems,Pimm_1984_ecosystems}, human infrastructures~\cite{Yang_2017_powergrids, Doyle_2005_internet} or trade networks~\cite{Grassia_2022_food_trade}.

In transport networks, perturbations often originate from node(s) or link(s) removal (complete failure) or congestion (partial failure) and take the form of delays, which generation and propagation depend on both local dynamics and network topology~\cite{hens2019spatiotemporal, chen2023finding}.
Most existing approaches to analyze such systems are data-driven, treating the generation and absorption of delays as processes that are independent of the propagation mechanism~\cite{pyrgiotis2013modelling, fleurquin2014characterization}.
This limits their ability to capture feedback effects and the coupling between structure and dynamics.
In public transport systems, a variety of endogenous and exogenous factors can cause delays at a location.
Air travel~\cite{Lacasa_2009,Fleurquin_2013,pyrgiotis2013modelling,Wu_2019} is susceptible to delays caused by weather, technical problems, or flight control failures.
Delays can quickly propagate to neighboring areas through network connectivity, affecting not only close, but also distant passengers through subsequent connections.
Delays can arise during dwell or running times from safety and operational constraints that might apply along the links, such as \emph{blocking sections} in railway systems, or on the nodes, such as traffic control and stuffing shortages in flight networks.
Although many studies have modeled delay propagation, for example using the max-plus algebra~\cite{Goverde_2007maxplus} or stochastic modeling~\cite{Meester_2007, Buker_2012_delayprop_largenerworks, Zieger_2018_buffertime,Buchel_2020_empirical_dynamics}, the influence of external stressors, such as heavy precipitations or strong winds, is typically not embedded into the model itself, but is treated rather as an input to the propagation mechanism~\cite{artime2024robustness}.
Climate variables have been demonstrated to produce disruption and breakdowns in transport systems such as the heavy rains of 2021 over Germany and the Benelux, and in 2012 in the UK~\cite{Jaroszweski_2014_uk2012, Szymczak_2025_german_rainfall}.

Several studies have successfully characterized the full or partial degradation of the transportation network through topological analysis.
For instance, passageways between cities have been utilized to assess network reliability~\cite{Ip_2011}.
Network centralities have been linked to network performance~\cite{Zhang_2015} and resilience to natural or man-made attacks~\cite{Bhatia_2015_railwayResilience}.
Optimization techniques for mitigating degraded states~\cite{Gedik_2014} highlight the importance of having multiple viable alternative routes, while rerouting heuristics and cost analysis help identify element criticality~\cite{Khaled_2015}.
Multi-agent modeling is at the basis of vulnerability and capacity reduction analysis~\cite{Cats_2016_partialcapacitydegradation, Malandri_2018, Massobrio_2024}, allowing model validation that replicates real-world delay distributions~\cite{Monechi_2018}.
Additional vulnerability assessments are conducted through the minimization of the cost of rolling stock rerouting~\cite{Bababeik_2017_multilink_blockage}.
Clustering of states in the principal component subspace can offer early warnings~\cite{Dekker_2019}, validated by multi-agent simulations.
Delays propagate along links and are amplified in constraints defined in several layers of connectivity (lines, trains, rolling stocks, and crews)~\cite{Dekker_2021_cascading, Daniotti_2024_delay_cascading_mitigation}.
We refer to~\cite{Faturechi_2015_transport_review, Gu_2020} for reviews on this topic.

Here we propose a minimal yet physically motivated model of perturbation propagation that explicitly couples the internal dynamics of the transport network with external climatic variables.
Specifically, this unifying approach takes into account the geographic embedding of the transport network and its exposure to external stressors, typically represented by climatic extreme events, while disregarding the complexities of the physical system (crew dispatch, rolling stock routing, etc.).
Instead of percolation analysis or cascade description due to attacks to the topology, the external stressor acts as a field that drives the dynamics out of the functional state.
Here, we consider the \ac{irn} and the \ac{crn} as special showcases that reproduce the delay generated by precipitations on rainy and extremely rainy days.
Finally, we analyze the inherent weaknesses of the coupled systems through the study of the model under global and local perturbation, and the social cost resulting from these functional disruptions.

Our results provide a unified perspective on the co-evolution of infrastructure functionality and environmental stressors within a simple framework.
Constrained by the focus on the latter, this framework can provide a more transparent view on perturbation generation.
The proposed propagation mechanism relies on a relatively simple linear diffusion model which is a well-established methodology in network science, but simplifies the real dynamics which are shaped by timetables, dispatching rules, bottlenecks and other details that all generate nonlinear operational responses.
Our goal is to develop a stylized risk indicator that provides a high-level understanding of the impact of an external stressor and not to give a detailed transportation engineering analysis.
Nevertheless, effective mitigation strategies, such as real-time monitoring systems and adaptive scheduling algorithms, can be developed to quickly respond to emerging perturbations beyond the specific application considered here.
For example, by identifying critical nodes or pathways within a network, stakeholders can implement targeted interventions to strengthen those areas, thereby reducing the likelihood of widespread disruptions.
In addition, the inclusion of the climatic stressor in the propagation model allows one to assess the robustness of the system in a wide range of what-if scenarios, a crucial feature for policy and decision-making.

Finally, our model's predictions suggest the presence of a global disruption phase of the transportation system at high levels of stress, and geographically localized disturbances at lower levels.
Our analyzes are particularly relevant as the frequency and intensity of extreme climate events are projected to increase in the coming decades~\cite{fischer2021increasing,rahmstorf2011increase}.
Attribution studies indicate that anthropogenic climate change has already increased the probability of high-precipitation extremes~\cite{risser2017attributable}.
In a world increasingly threatened by climate change~\cite{steffen2018trajectories}, understanding how these events interact with critical infrastructure is essential to anticipate systemic risks and develop evidence-based resilience strategies.

\section{Results}
\subsection{Perturbation transportation model}

The functionality of a transport network can be influenced and degraded by the interaction of its topology or dynamics with external stressors.
The latter induces local or global perturbations in the system dynamics that propagate throughout the connectivity pattern of the network.
If left unregulated, perturbations can completely disrupt the system functionality, affecting the people who rely on such an infrastructure.

\begin{figure}[!ht]
	\includegraphics[width=\linewidth]{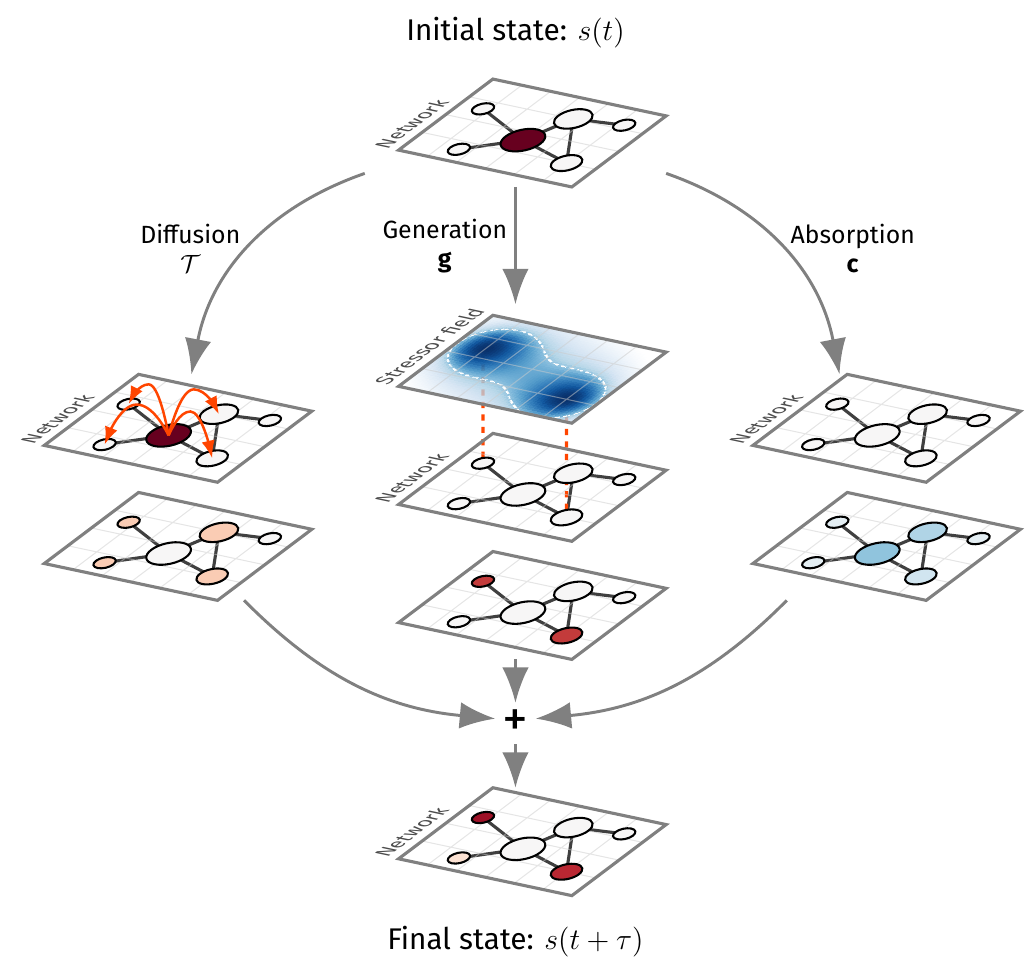}
	\caption{
		\textbf{Illustrative representation of the perturbation propagation model.}
		The model described by Eq.~\eqref{eq:generic_model} is composed by three independent processes that operate at each time-step $t$: the \emph{diffusion} process (left) that enable perturbation propagation, the \emph{generation} process (center) induced by the external stressor field, and the \emph{absorption} process (right) regulated by the system capacity to return to a functional state.
		Blue to red colors represent negative to positive perturbations.
	}
	\label{fig:model_example}
\end{figure}

To formalize this, we disregard the microscopic physical mechanics of the system of interest, but learn the response of its building blocks to the external stressor.
Consider a network $G(V,E)$ determined by a set of $N_V$ nodes ($V$) and a set of $N_E$ edges ($E$) connecting these nodes.
The nodes and edges of this network are embedded in a space where the external stressor field $\phi(x, t)$ is defined.
The functional activity of the system can be continuously monitored by the state of the externally induced perturbation $\mathbf s(t)=\{s_1(t),\ldots, s_{N_V}(t)\}$, where $s_i(t)$ is the amount of perturbation seen at the node $i$ at time $t$.
The evolution of the perturbation can be described as follows:
\begin{equation}
	\mathbf s(t+\tau)  = \alpha\mathcal T(t) \mathbf s(t) + \beta \mathbf g(t) -\gamma \mathbf c(t),
	\label{eq:generic_model}
\end{equation}
where $\tau$ represents the smallest temporal step in which the discrete dynamics are updated.
Negative values are set to zero because we assume that negative delays are meaningless in this context.
The evolution of the perturbation is controlled by three fundamental processes corresponding to each term of Eq.~\ref{eq:generic_model} and depicted in Figure~\ref{fig:model_example} for a visual representation.
The first term represents a diffusion process that allows the perturbation from each node of the network to propagate to its neighbors.
This term is described by the transition matrix $\mathcal T(t)$, whose elements $\mathcal T_{ij}(t)= p_t(i|j)$ represent the probability that, from node $i$, the perturbation will reach node $j$.
Note that, at this stage, the transition matrix should be considered time-dependent, reflecting the temporal evolution of the network of interest.
The generation process is regulated by the delay generation rate $\beta g_i(t)$, where $g_i(t)$ depends on the intensity of the stressor at each node $i$ or along its incident links.
Lastly, the absorption process is controlled by the delay absorption rate $\gamma c_i(t)$, where $c_i(t)$ is the maximum capacity of the node $i$.
For maximum flexibility, the model is characterized by three parameters: $\alpha$ representing the self-interaction of the dynamical process along an edge, $\beta$ representing the perturbation creation rate, and $\gamma$ representing the absorption rate.

The interaction with the external stressor field $\phi(x,t)$ is the driver of the perturbation generation process described by $\mathbf g(t)$ in Eq.~\ref{eq:generic_model}.
If the interaction occurs at the node location, the process follows $\mathbf g_i(t) = f(\phi(x_i, t))$ where $x_i$ is the location of the node $i$ in the domain of the external stressor field.
On the other hand, if the interaction occurs along the edges of the network, the term $\mathbf g_i$ at node $i$ will depend on the values of the external field along all edges incident to that node.

\subsection{Delay propagation on railway networks due to rainfalls}

Although the above model is general to a number of transport phenomena, we will focus on a railway network where the perturbations are characterized by the delays that propagate on the system.
For the sake of simplicity, we only consider delayed trains, neglecting train cancelations or rerouting.
Within this framework, delays can propagate as a cascading mechanism seeding primarily along the lines where infrastructure limitations, safety guidelines, and external factors interplay with running and dwell times.
Let us suppose that part of the delays experimented on a railway system may be attributed to the weather conditions, in particular to the rainfalls.
For instance, operational limitations are put in place during extreme events to enhance the transportation safety~\cite{Sun_2022_velocity_reduction_railways}, decreasing the transport capacity of railway lines and injecting delays in the transport system.
The generation process acts along the line and can be written in the following way:
\begin{equation}
	\mathbf g_i(t) = \sum_j \mathcal W_{ij}(t)\int_{x_j}^{x_i} \phi(x, t)dx,
	\label{eq:line-integral}
\end{equation}
where $\mathcal W_{ij}(t)$ is the number of trains traversing the link between node $i$ and node $j$ at the given time $t$, and the integral is carried out along the rail line.
The absorption mechanism is implemented at nodes where scheduled buffer times are designed to mitigate accumulated delays.
We consider a time-independent node capacity corresponding to the maximum throughput of each node $\mathbf c_i(t) =\max_t \sum_j \mathcal W_{ij}(t)$.

In this framework, $\alpha$ represents the self interaction of trains along the same line and accounts for the additional number of minutes of delays each train induces on trains following the same line.
The following parameter, $\beta$, corresponds to the delay generation rate per unit line length and amount of rain, while $\gamma$ represents the amount of minutes of delay absorbed in each station per station capacity.
Finally, we take the basic update step $\tau$ to be of one hour.

\subsection{Optimization and validation}

We optimize the prediction of this model by comparing them with the delays reported by the railway operator.
Since train delays can arise from a variety of reasons, we consider only the excess of reported delays with respect to the expected delay, which is chosen to be the median of the historical delay at each station.

\begin{figure*}[t!]
	\begin{center}
		\includegraphics[width=0.95\textwidth]{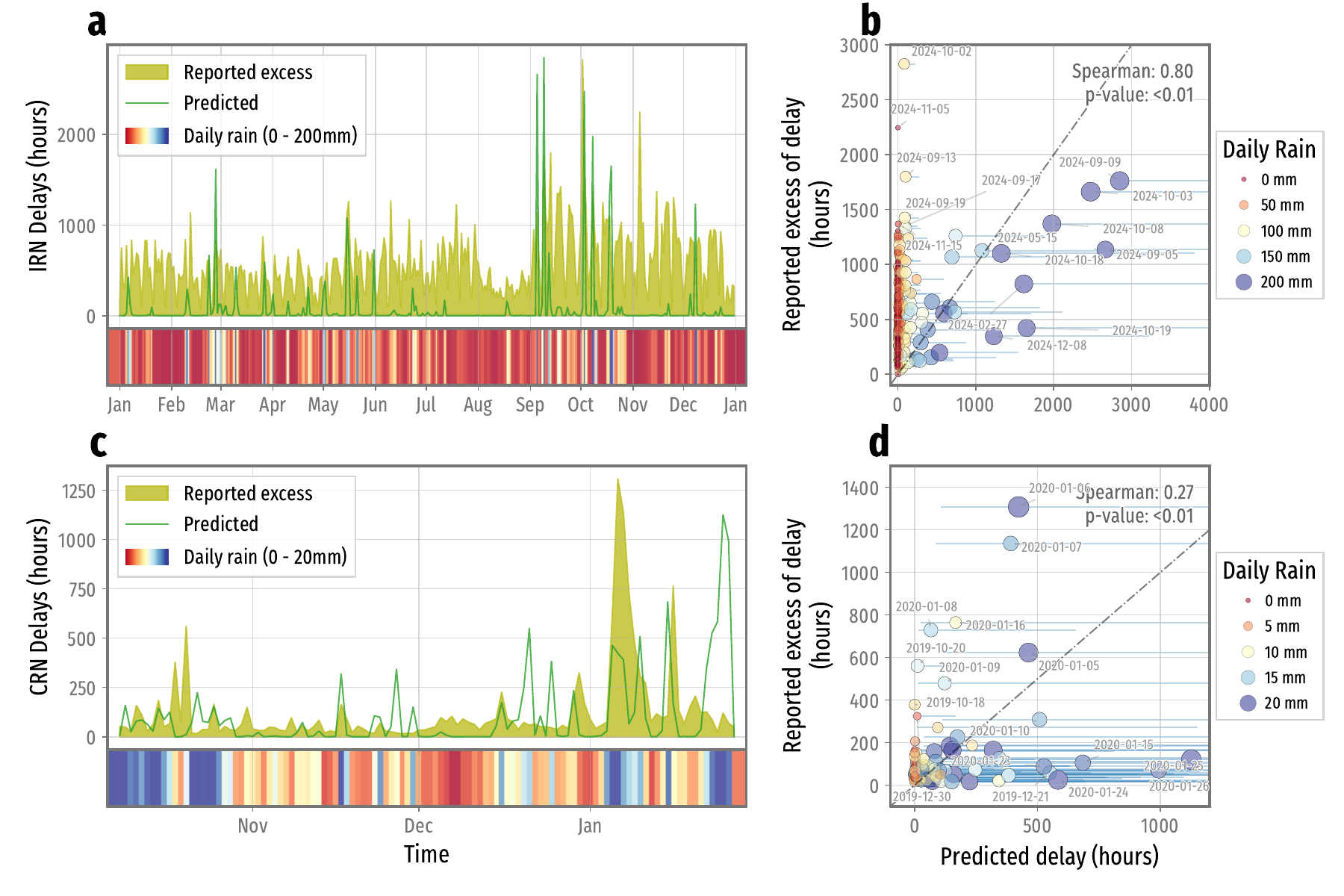}
	\end{center}
	\caption{
		\textbf{Daily delay validation for 2024.}
		Time-series of rain, predicted delays and reported excess delays for \ac{irn} (\textbf{a}) and \ac{crn} (\textbf{c}).
		Comparison of predicted daily delays and reported excess delay for \ac{irn} (\textbf{b}) and for \ac{crn} (\textbf{d}).
		Color and size correspond to the cumulative amount of rain experienced by the railway system in each day.
		During early September and October, overlapping lines of panel \textbf{a} partially obscure the agreement, which becomes apparent in panel \textbf{b}.
		One can notice in \textbf{b} (\ac{irn}) how, in rainy days (blue dots), the predicted delay correlates to the reported excess of delay.
		In the high-speed Chinese railway system, rainfalls along the costal line do not affect the functionality of the service, producing the points in the lower part of panel (\textbf{d}).
		Horizontal blue lines represent the confidence interval of each point.
	}\label{fig:validation}
\end{figure*}

\begin{table}[t!]
	\begin{center}
		\label{table:params}
		\begin{tabular}{ccr@{--\,}lcr@{--\,}l}
			\toprule
			          & \multicolumn{3}{c}{\ac{irn}} & \multicolumn{3}{c}{\ac{crn}}                                                  \\
			Parameter & value                        & \multicolumn{2}{c}{CI}       & value & \multicolumn{2}{c}{CI}                 \\
			\midrule
			$\alpha$  & 1.0005                       & 1.0                          & 1.007 & 1.000                  & 1.0   & 1.019 \\
			$\beta$   & 7.86                         & 6.24                         & 15.27 & 3.00                   & 0.61  & 32.94 \\
			$\gamma$  & 2.20                         & 1.43                         & 3.63  & 0.033                  & 0.007 & 0.121 \\
			\bottomrule
		\end{tabular}
	\end{center}
	\caption{ Model parameters fitted to the learning sets.
		While the values of the diffusion parameter ($\alpha$) are compatible, $\beta$ and $\gamma$, which account for the system response to the stressor, are dataset specific.
		Although larger confidence intervals in the \ac{crn} are due to the smaller dataset size, incorporating geographic dependence of parameters may improve this by accounting for heterogeneous system sensitivity to the external stressor (see Supplementary Information).
	}
\end{table}

We divide the \ac{irn} dataset into a learning set (from 2021-2023) and a test set (data from 2024) and perform a $k$-fold cross-validation to optimize the model parameters (with $k=4$).
The learning phase (see Methods for details) optimizes the model parameters and yields $\alpha=1.0005$, $\beta=7.86$ and $\gamma=2.20$ (see Table~\ref{table:params} for the confidence intervals and the corresponding values for the dataset \ac{crn} dataset).
Figure~\ref{fig:validation} shows how the predicted delay from the real stressor data compares with the reported excess delay.
From the comparison between the predicted and reported delays, a natural split of the results arises.
In rainy days (blue dots) where the daily average rain exceeds \qty{144}{mm} (or \qty{6}{mm.h^{-1}}), the predicted delay correlates with the reported excess of delay, with a Spearman's coefficient of about 0.80 (p-value $<0.01$).
The reader should keep in mind that $\beta$ and $\gamma$ represent competing processes (such as creation and absorption of perturbations) and across optimization steps, they can move together showing a strong positive correlation.

On the 19 rainiest days of the \ac{irn}, we are able to differentiate between rainfalls that produce a wide system disruption and those that do not.
The latter are characterized by an external stressor distributed or concentrated above low susceptibility regions (see Supplementary Information for additional examples).
During less rainy days, when the external stressor field is too weak or absent, we are unable to predict any delay, since the driving cause of their generation is independent of the climate variables.

The \ac{crn} dataset is smaller compared to the \ac{irn}, containing only 112 days of reported delays restricted to high-speed trains and concentrated in winter, with less intense rains.
Due to these reasons, our model displays a lower predictive power with a weaker correlation (see Figure~\ref{fig:validation}).
Nevertheless, it can reproduce the network distress due to abundant precipitations in a few days, and in particular on January 6, 2020 when abundant snowfalls disrupted flight and train connections~\cite{snowarticle_url}.
In the lower part of Figure~\ref{fig:validation}d, for numerous days, the model overpredicts the delays caused by the rain.
Those days are characterized by rainfall concentrated in the southern part of the country, which stations are less sensitive to the external stressor (see the Supplementary Information for a discussion).
The geographical heterogeneity of the model performance highlights how a large system, such as the \ac{crn}, might have an inhomogeneous response to the external stressor and a spatially dependent $\beta$ and $\gamma$ might be introduced.
In particular those parameters might be spatially distributed as $\beta(x)$ and $\gamma(x)$, or constant in smaller sub-regions, with the downside of parameter proliferation.

This approach outperforms a simple prediction based on the optimization of a Ridge regressor on the same learning and testing datasets.
The latter provides a Spearman's coefficient of only 0.46 (p-value 0.042) in the case of \ac{irn} and 0.20 (p-value 0.046) in the case of \ac{crn} (see Supplementary Information).

\begin{figure*}[t!]
	\begin{center}
		\includegraphics[width=0.9\textwidth]{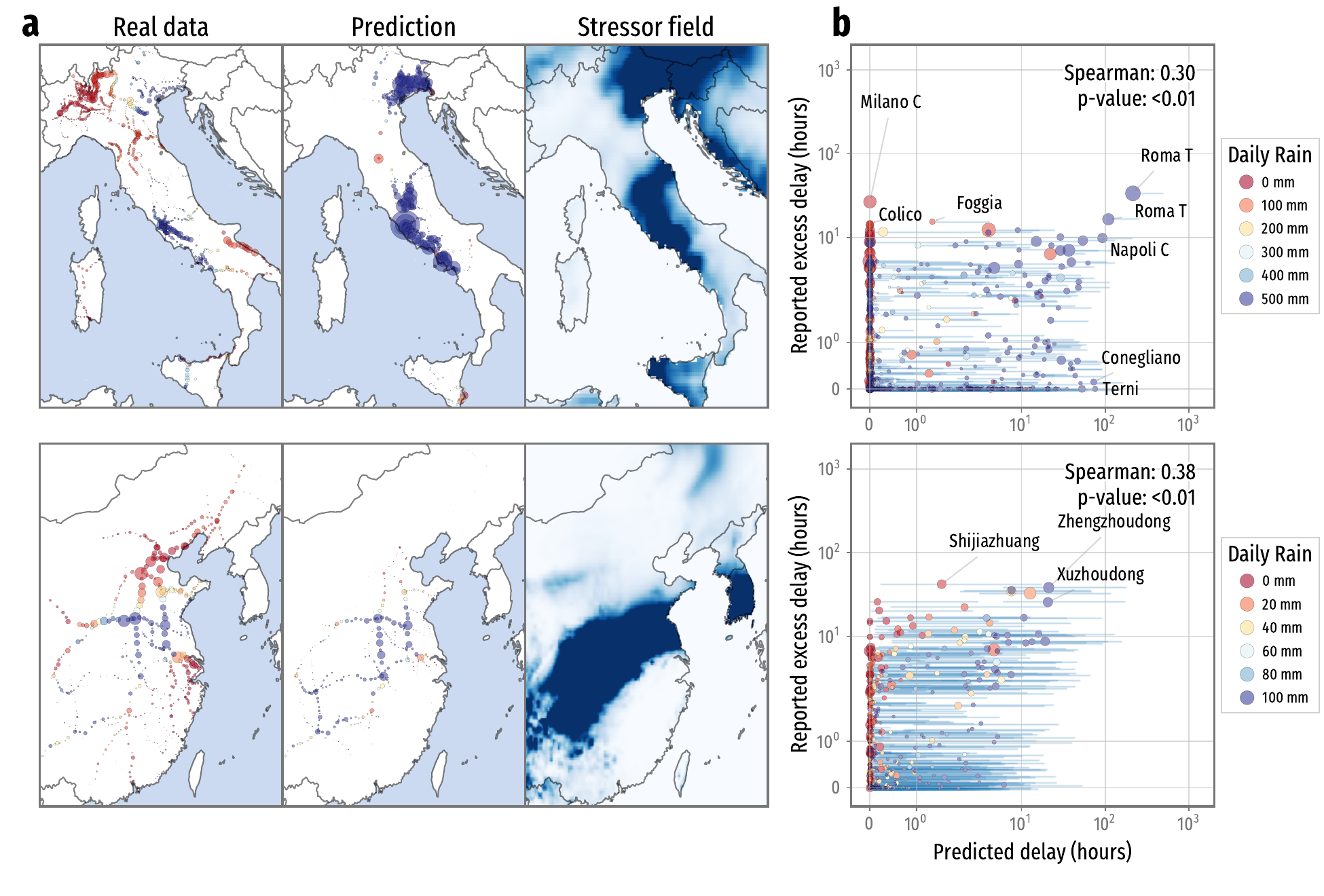}
	\end{center}
	\caption{
		\textbf{Local daily delay validation in rainy days.}
		Two rainy days: 9th of Sept. 2024 from \ac{irn} (above) and 6th of Jan. 2020 from \ac{crn} (below).
		Notice how nodes with predicted delays (\textbf{a} center) concentrate in and around the areas with strong external stressor field (\textbf{a} right).
		Node size is proportional to the real or predicted delay (\textbf{a}) or to the capacity (\textbf{b}).
		The Spearman's coefficient is computed on stations with nonzero delay.
		Horizontal blue lines represent the confidence interval of each point.
	}\label{fig:validate_day}
\end{figure*}

At a granular level, it is interesting to compare the predicted delay at each station with the corresponding reported excess delay.
Figure~\ref{fig:validate_day} reports this comparison for the one of the rainiest days for each dataset, with a similar but weaker result than Figure~\ref{fig:validation}.
In stations affected by a strong external stressor field, the predicted delay tends to overestimate the excess of reported delay to compensate for stations located in non-rainy areas, whose delays are most probably unrelated to the climate stressor.

\subsection{Global system response}

Under continuous stress, the system can reach a steady state characterized or not by an endemic delay.
Suppose a time-independent external stressor field $\phi(x,t)=\qty{10}{mm.h^{-1}}$ of rainfall and a transfer matrix $\mathcal T (t)= \mathcal T $ corresponding to a busy hour (e.g. 8AM of a working day).

\begin{figure*}[t!]
	\begin{center}
		\includegraphics[width=\textwidth]{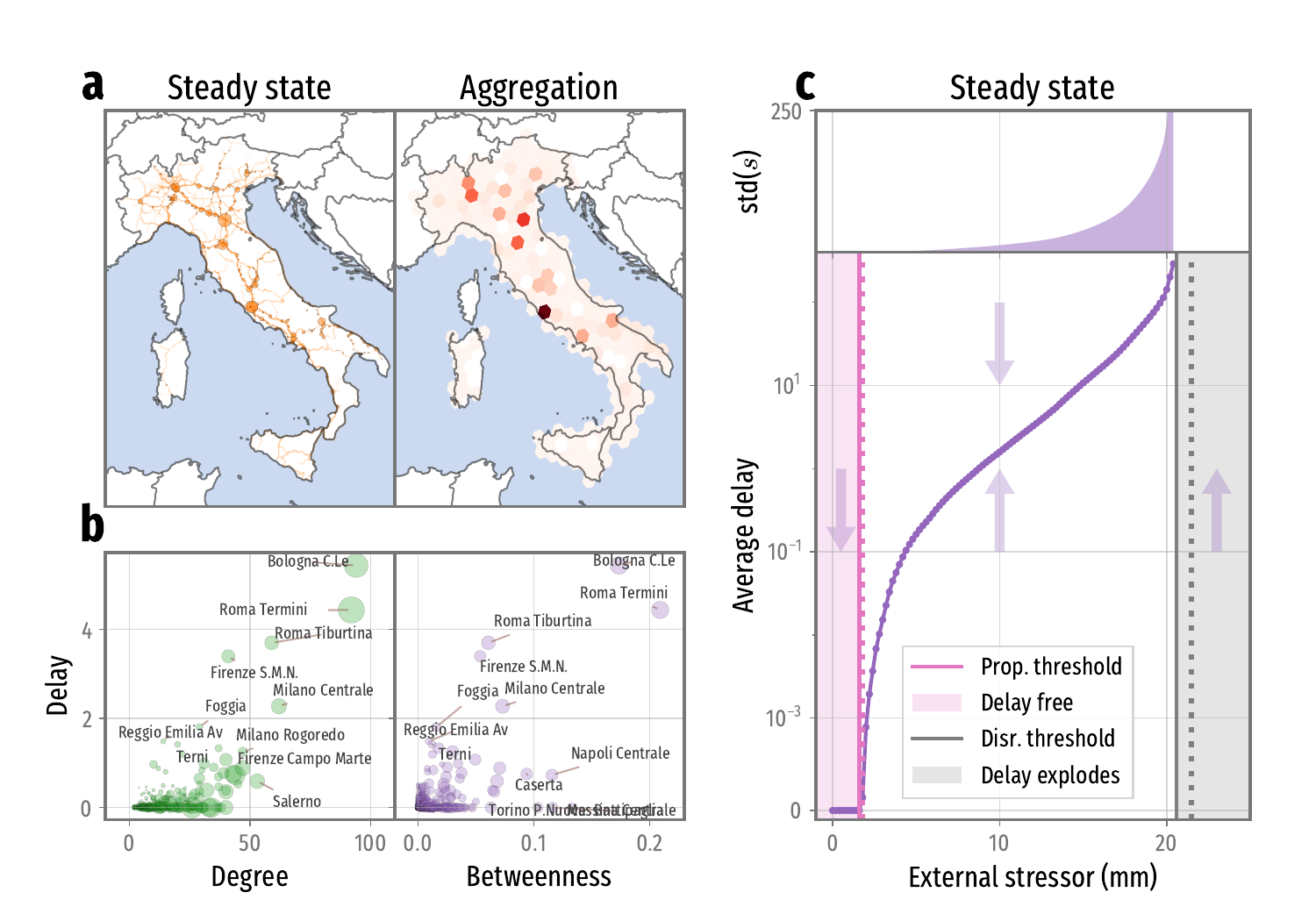}
	\end{center}
	\caption{
		\textbf{Steady state distribution of the model under a constant external stressor field.}
		Distribution of delay (size) on the map and its aggregation on a hexagonal lattice (\textbf{a}) with an external stressor field of 10 mm of hourly rain.
		The weak correlation of the steady state with degree distribution and betweenness centrality (\textbf{b}) shows the shortcoming of topology only indicators.
		The steady state phase diagram (\textbf{c}) includes three phases marked by color codes (no delay (pink), endemic finite state attractor (white), delay diverges to infinity (gray)).
		Phases are separated by the propagation and the disruption thresholds (dashed lines correspond to the approximate theoretical calculation).
		The standard deviation $\text{std}(s)$ is reported on top.
	}\label{fig:steadystate}
\end{figure*}

Figure~\ref{fig:steadystate} displays the delay distribution reached by the model after the initial transient period.
The final distribution correlates with topological quantities such as degree (Spearman's coefficient 0.62, p-value $< 0.01$ for \ac{irn}, and 0.42, p-value $<0.01$ for \ac{crn}) and weighted betweenness centrality (Spearman's coefficient 0.63, p-value $<0.01$ for \ac{irn}, and 0.24, p-value $<0.01$ for \ac{crn}), although the topological feature is not enough to completely determine the system behavior.

The steady-state phase diagram (Figure~\ref{fig:steadystate}c) shows three phases at different values of the external stressor field separated by two thresholds.
The first phase is characterized by an absorption process that outperforms the external stressor field generation process, resulting in no delays.
Beyond the \emph{propagation threshold}, the second phase is characterized by a finite long-term average delay, resulting in a dynamical equilibrium of endemic delay.
The third phase, above the \emph{disruption threshold}, is characterized by a constant growth rate of delays that diverge over time, resulting in systemic collapse, here represented by the global jamming of the entire system.

The approximate values for the propagation and disruption thresholds can be analytically computed.
The propagation threshold is characterized by an external stressor field such that $\beta \mathbf g_i(\phi) < \gamma \mathbf c_i \ \forall i$, below which the perturbation is completely absorbed by the system.
Above this threshold, in at least one node, the generation exceeds the absorption.
The higher disruption threshold is found at $\beta \langle \mathbf g\rangle = \gamma \langle \mathbf c\rangle$ where, on average, the global perturbation generation exceeds global absorption.
Figure~\ref{fig:steadystate}c displays the approximate theoretical values as dashed lines and the numerical values as solid lines (see Table~\ref{table:thresholds} for the values).
The steady state phase diagram of the \ac{crn} displays the same behavior (see the Supplementary Information for details).

A signature of the lower propagation threshold can be found in the Supplementary Information, where the excess of reported delays and the number of cancelled trains increase, on average, when the rainfalls reach a minimal intensity.
The conditions necessary to reach the disruption phase are not met in our datasets, but, while both thresholds depend on the network characteristics, disruption and breakouts have been found in the literature.
In the heavy rains of July 2021 in Germany and Benelux, total precipitations of up to \qty{100}{mm} per day affected the functionality of the German-wide network~\cite{Szymczak_2025_german_rainfall}.
During the 2012 storm that affected the UK railroads, precipitations reaching up to \qty{40}{mm.h^{-1}} heavily disrupted the rail transportation service, with cumulative delays of about 10000 minutes~\cite{Jaroszweski_2014_uk2012}.

\begin{table}[!htb]
	\begin{center}
		\label{table:thresholds}
		\begin{tabular}{cccr@{--\,}lc}
			\toprule
			Network & Threshold      & Analytical       & \multicolumn{2}{c}{CI} & Numerical                    \\
			        &                & \unit{mm.h^{-1}} &                        &           & \unit{mm.h^{-1}} \\
			\midrule
			\multirow{2}{*}{\ac{irn}}
			        & Propagation T. & 1.76             & 1.44                   & 1.82      & 1.60             \\
			        & Disruption T.  & 21.47            & 17.61                  & 22.23     & 20.60            \\
			\midrule
			\multirow{2}{*}{\ac{crn}}
			        & Propagation T. & 0.34             & 0.21                   & 1.16      & 0.34             \\
			        & Disruption T.  & 1.87             & 0.60                   & 3.27      & 1.86             \\
			\bottomrule
		\end{tabular}
	\end{center}
	\caption{ Phase diagram thresholds.
		Propagation and disruption thresholds are reported for both \ac{irn} and \ac{crn}.
		Analytical results are reported with their confidence intervals and compared to the numerical results.}
\end{table}

Restricting the external field action to one full working day, see Supplementary Information for details, the resulting steady state is characterized by two phases separated by a continuous phase transition.
At low values of the external stressor field, the average delay is negligible and the absorption process is dominant.
Above the transition, the generation process dominates, and the delay increases linearly with the value of the external stressor field.

\subsection{Local system response}

An important application of this model is the detection of weak parts of the network, the first step towards weakness mitigation.

To this end, we introduce the node \emph{perturbability}, a new centrality that takes inspiration from network coverage~\cite{De_Domenico_2014_coverage}.
Perturbability accounts for the amount of delay that is generated in the system after a local perturbation acts for a short period of time on the surroundings of a given node.
In particular, a strong external stressor field is switched on for $\delta\tau=1$ hour, at \qty{100}{mm.h^{-1}} of rain, in the surrounding of a given node (in a circle of about \qty{50}{km}).
The perturbability is then described as the amount of delay generated on the system during the next 24 hours.
\begin{equation}
	p_i = \sum_{j=1}^{N_V} \sum_{t=1}^{24} s_j(t) dt
\end{equation}

\begin{figure*}[t!]
	\begin{center}
		\includegraphics[width=1\textwidth]{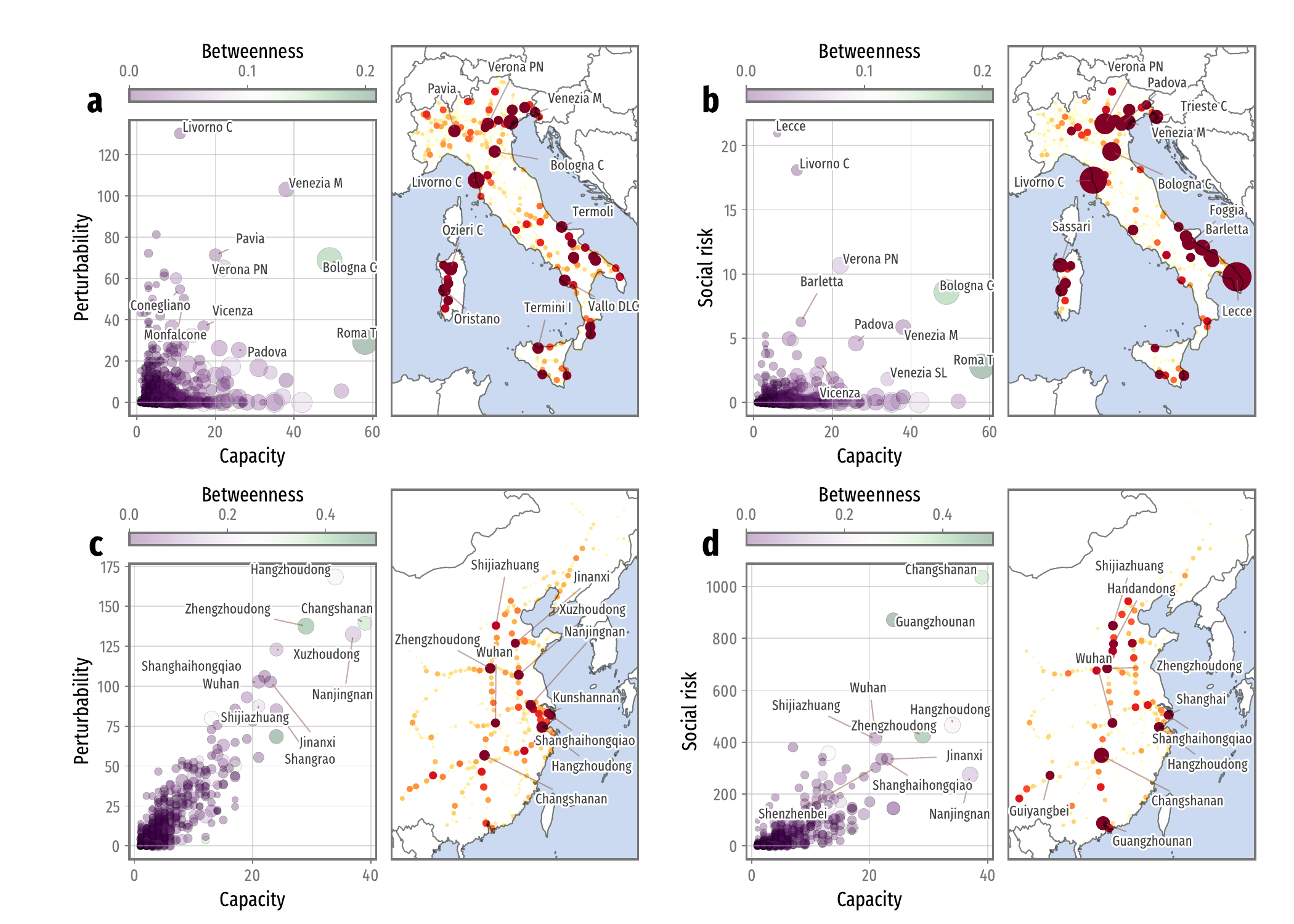}
	\end{center}
	\caption{
		\textbf{Perturbability of the system and social risk.}
		Each node is perturbed by a strong external stressor field for a short period of time and left to evolve according to the model for 24 hours.
		High perturbability appears on the east part of Italy (\textbf{a}) and is compared to node capacity. In \ac{crn} (panel \textbf{c}), pertubability is driven by the station centrality.
		Left panels display the distribution of social risk for \ac{irn} (\textbf{b}) and \ac{crn} (\textbf{d}) on the country and is compared to node capacity.
		Node size and color on the maps represent degree and betweenness centrality respectively.
	}\label{fig:coverage}
\end{figure*}

Figure~\ref{fig:coverage}a,c displays the geographic distribution of perturbability as well as its relation to station betweenness.
A noticeable difference between the two datasets arises, where in \ac{crn} perturbability is driven by the topological feature, while in \ac{irn} high perturbability stations are clustered on the east part of the Peninsula and on the islands.

Inspired by risk management~\cite{Aven_2016_risk_assessment}, we devise a measure of the social cost due to the weakness of the station.
Using the estimated population density of the WorldPop project~\cite{Worldpop}, we define as \emph{station risk} the product of the delays generated and the affected population (social cost) at each station:
\begin{equation}
	r_i = \sum_{j=1}^{N_V}\sum_{t=1}^{24} \text{pop}_j s_j(t) dt\,,
\end{equation}
where $\text{pop}_i$ represents the population affected by station $i$.

The resulting risk distribution is depicted in Figure~\ref{fig:coverage}b,d.
Despite showing similar results to the above perturbability analysis, the risk distribution highlights in both networks how the population distribution can change how disruption is perceived.
The stations of larger cities, such as Lecce and Verona in \ac{irn} and Guangzhounan and Shangai in \ac{crn}, increase their rank due to the combination of high population and high perturbability.

\section{Discussion}

Critical infrastructures are constantly under stress due to the impact of external events, including those generated by climate.
Understanding how these impacts affect the functionality of the system becomes of utmost importance to put in place recovery and mitigation strategies.
This becomes even more central in a changing climate.

In this study, we have developed a novel perturbation propagation model with simple building blocks.
This describes the generation, propagation, and absorption of perturbations across those infrastructures, specifically focusing on the impact of weather variables.
We selected two national railway systems under the influence of rainfall as a specific case study.
Our findings indicate that as the intensity of the rainfall increases, the resulting perturbations within the network also increase displaying three phases: complete absorption, endemic perturbations, and system disruption.
The model shows that, under certain conditions, these delays can exhibit unbounded growth, suggesting critical vulnerabilities in the current infrastructure.
Although unbounded growth conditions are never matched by the weather variable in the data, signature events have been observed recently~\cite{Szymczak_2025_german_rainfall, Jaroszweski_2014_uk2012}.
Furthermore, local analysis unveils areas with higher sensitivity to external factors that better emerge under a risk assessment where the social cost of emerging delays is the population affected by the disruption.

As weather patterns become increasingly erratic and extreme due to climate change, the frequency and intensity of rainfall events are expected to change over the next decades.
This trend poses a substantial risk to the resilience and efficiency of train networks, which are critical to public transport and freight movement.
Our model introduces a novel tool for understanding these dynamics and can aid in the development of more resilient transportation systems.
It helps identify critical rainfall thresholds that lead to significant delays or stations with a higher capacity to influence the system.
The stakeholders can then prioritize upgrades to vulnerable sections of the network, implement real-time monitoring systems, and develop contingency plans to minimize disruptions.

Our study still displays some weak points that can be improved in future work.
Firstly, a national railway system is not isolated from the surrounding networks (e.g., other national systems or other transportation layers, such as air traffic), and delays might propagate between them.
To this end, a multilayer perspective~\cite{mishra2012performance,de2016physics,de2023more} is necessary to inform decisions with greater precision.
Secondly, additional sources of delays, such as fields beyond precipitation (e.g., heatwaves, or extreme wind), might be included in the model to provide a comprehensive description, without changing the underlying mathematical framework.
Finally, a complete description of the system, including train cancellation, dispatch rules, or constraints, can improve our understanding of the microscopic dynamics.
In the current manuscript we are interested in the minimal description of the interaction of the system with an external climate variable.
To this end, we disregard those complex mechanisms and consider the linear approximation of the system.
In the next future, such enhancements might even provide a fertile ground for the development of dedicated digital twins to assess the resilience of complex human infrastructures.

In conclusion, as climate change continues to influence weather patterns, the need for robust models that can accurately predict the impacts of environmental variables on transportation networks becomes increasingly urgent.
Future research should focus on further refining the model and exploring its applicability to other transportation systems, thereby broadening our understanding of perturbation propagation in the context of climate variability.

\section{Materials and Methods}

\subsection{Data}

\subsubsection{Railway networks}

For both railway networks, the station location and the line shapes have been extracted from the OpenStreetMap project and merged with the data provided by TrainStats (\url{https://trainstats.altervista.org/}) for \ac{irn} and \cite{Zhang_a_2022} for \ac{crn}.
From the schedule data, we computed the median delay observed at each station during one hour and its capacity.
The latter corresponds to the maximum train throughput observed in the given station for one hour.
From WorldPop~\cite{Worldpop} data, we assign the population served by each station, tessellating the country surface in Voronoi cells.
For each edge, we extracted the line shape along which we perform the line integral of Eq.~\ref{eq:line-integral}.

Consider $\mathcal W_{ij}(t)$ as the average number of trains that arrived at node $i$ at a time $[t;t+\tau]$ and originated at node $j$.
We infer the values of the transition matrix $\mathcal T$ as:
\begin{equation*}
	\mathcal T_{ij}(t) = \frac{\mathcal W_{ij}(t)}{\sum_i \mathcal W_{ij}(t)}
\end{equation*}
Here we consider $\mathcal W$ and $\mathcal T$ to depend on the hour of the day, the day of the week (if the day of interest is Sunday or not), and the month (if the month of interest is August or not).

\subsubsection{External stressor field}
The total precipitation dataset was collected from the ECMWF reanalysis (ERA5)~\cite{munoz2019era5} and has been retrieved from the Copernicus project using the \texttt{Python} APIs.

\subsection{Optimization of the model parameters.}

We fit the model parameters ($\alpha$, $\beta$ and $\gamma$) comparing the predicted delay to the excess delay reported by the railway operator.
By excess delay $s_i^e(t)$, we refer to the amount of delay beyond the expected value at each station, which we define as the median delay observed at each station:
\begin{equation*}
	s_i^e(t) = \max (s_i^\text{real}(t) - \tilde s_i, 0)
\end{equation*}
where $s_i^\text{real}(t)$ is the real delay observed at station $i$ during day $t$.

Let us define $s'_i(t)$ the delay cumulated at the station $i$ throughout the day $t$, as a result of the simulations of the model.
We use as objective function (to be minimized) the following:
\begin{equation}
	f = \sum_{i,t} \left(s^e_i(t) - s'_i(t)\right)^2
\end{equation}

The minimization of the above objective function is performed through a Nelder-Mead bounded algorithm, with parameter bounds: $\alpha \in [1.0, \infty]$, $\beta\in[0,\infty]$, and $\gamma\in[0,\infty]$).

\subsection{Steady state theoretical thresholds}

\paragraph*{Propagation threshold}
Let us suppose starting from an initial state where there are no delays in the network.
For the external stressor field to be able to generate some perturbation in the system, the generation process should outperform the absorption process.
This is accomplished if at least for one node $\beta g_i > \gamma c_i$.
Below this value, the perturbation will be completely absorbed by the system.

\paragraph*{Disruption threshold}
Consider the order parameter $\langle\mathbf s(t)\rangle$ that represents the average perturbation at time $t$ across all nodes.
Its value is forced to increase steadily if
\begin{equation*}
	\langle\mathbf s(t+1) \rangle = \langle \mathcal T\mathbf s(t)+\beta\mathbf g-\gamma\mathbf c\rangle > \langle\mathbf s(t)\rangle.
\end{equation*}
Since $\mathcal T$ is supposed to be a stochastic matrix, the $l_1$ norm of the vector $\mathbf s(t)$ is preserved and:
\begin{equation*}
	\langle\mathbf s(t+1) \rangle = \langle \mathbf s(t)\rangle+\beta\langle\mathbf g\rangle-\gamma\langle\mathbf c\rangle.
\end{equation*}
Finally, the order parameter will continue to grow if $\beta\langle\mathbf g\rangle>\gamma\langle\mathbf c\rangle$.

\paragraph{Acknowledgments.}

Data and code associated with this study is available at \url{https://github.com/CoMuNeLab/adaptsys_train-delays} for the \ac{irn} dataset, and at \url{https://github.com/CoMuNeLab/adaptsys_chinese_train_delays} for the \ac{crn} dataset.

MF, TS, and MDD acknowledge financial support from MUR (Ministero dell'Università e della Ricerca) funding within the FIS (Fondo Italiano per la Scienza) (DD n. 1219 31-07-2023) project number FIS00000158 (CUP C53C23000660001).
MF also acknowledges the partial financial support from the INFN (Istituto Nationale di Fisica Nucleare) LINCOLN grant.

\bibliographystyle{science}
\bibliography{biblio}

\pagebreak
\appendix
\renewcommand{\thefigure}{S\arabic{figure}}
\setcounter{figure}{0}
\section{Supplementary Information}
\subsection{CRN rain/delay correlation}

\begin{figure}[!htb]
	\begin{center}
		\begin{tabular}{cc}
			IRN                                                                        & CRN \\
			\includegraphics[width=0.49\textwidth]{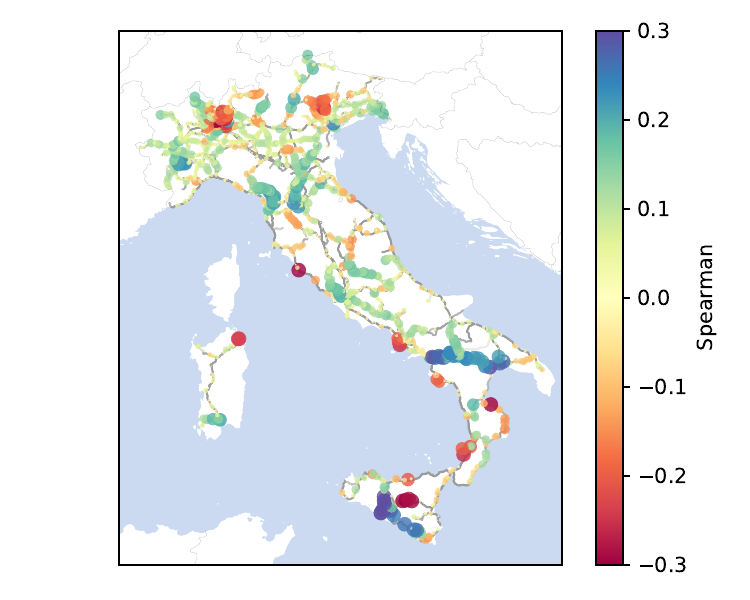} &
			\includegraphics[width=0.49\textwidth]{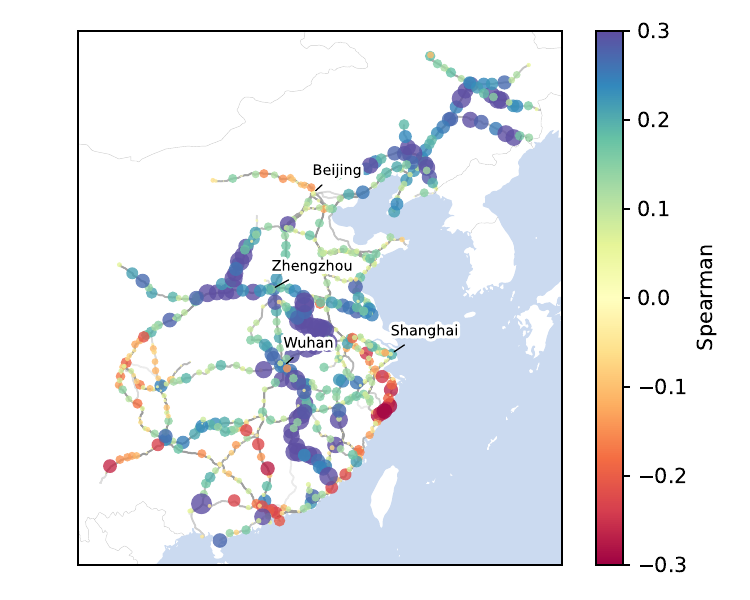}
		\end{tabular}
	\end{center}
	\caption{
		\textbf{Spearman correlation distribution}.
		Spearman correlation between excess of reported delay and rain at each train station.
		The central-northern region of CRN is particularly susceptible to perturbations driven by the rainfalls.
		Point size is proportional to $-\log(\text{pvalue})$.
	}\label{fig:crn_rain-delay}
\end{figure}

Figure~\ref{fig:crn_rain-delay} displays the correlation (Spearman's coefficient) between daily rainfalls and the excess of reported delay in IRN and CRN.
Notice how the stations susceptible to rainfalls (blue dots) are dispersed (with few local exceptions) in the IRC while they are concentrated on the central-northern area, in particular on the lines Shangai-Zhengzhou and Wuhan-Zhengzhou.

\newpage
\subsection{Rainy days with low level of delays (IRN)}

\begin{figure}[!ht]
	\begin{center}
		\includegraphics[width=0.8\textwidth]{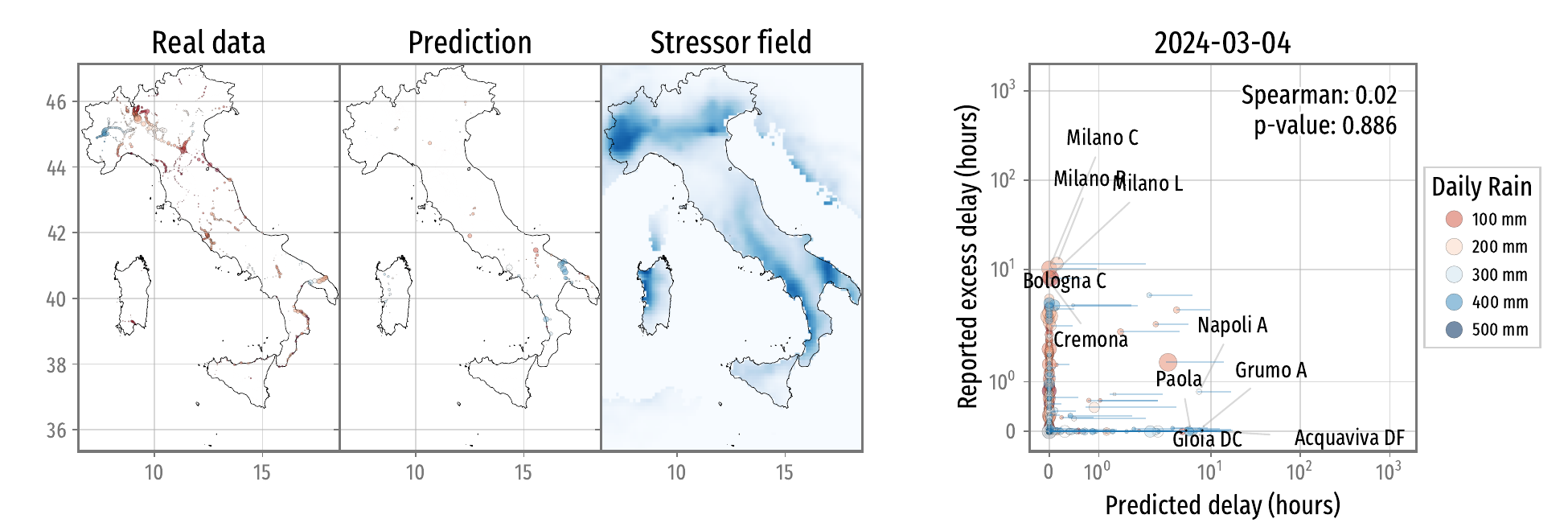}
		\includegraphics[width=0.8\textwidth]{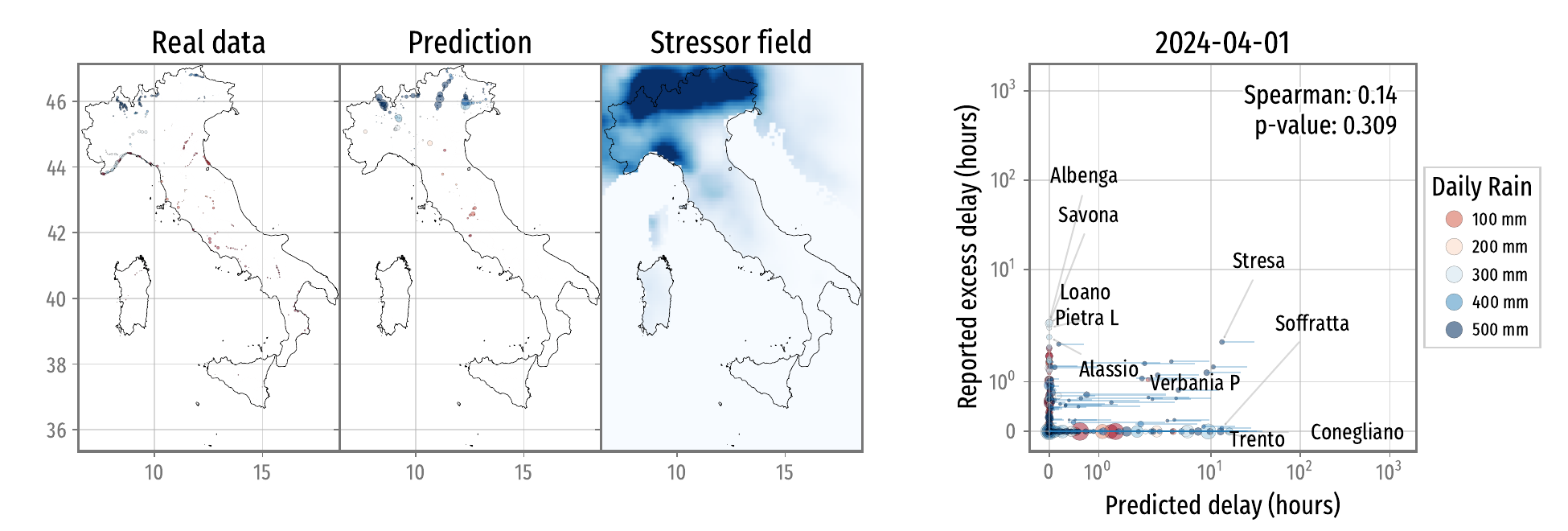}
	\end{center}
	\caption{Rainy days with low level of system disruption}\label{fig:lowdelays}
\end{figure}

Fig.~\ref{fig:lowdelays} displays two days with high precipitation and low predicted disruptions of the railway system.
On 4 March the rains were abundant but distributed over the whole country with local intensity too low to trigger perturbations; on 1 April the precipitations concentrated in locations with low perturbability.

\newpage
\subsection{Daily stressor intensity response}

Let us simulate the delay creation throughout a full working day with varying external stressor field intensity.
The cumulative daily delay driven by the external stressor field at each station increases with the strength of it.
The increasing delay converges to a linear growth for large values of the external stressor field as the second term of Eq. 1 in the main document becomes dominant in the dynamics.

\begin{figure}[htb]
	\begin{center}
		\begin{tabular}{cc}
			IRN                                                                       & CRN \\
			\includegraphics[width=0.49\textwidth]{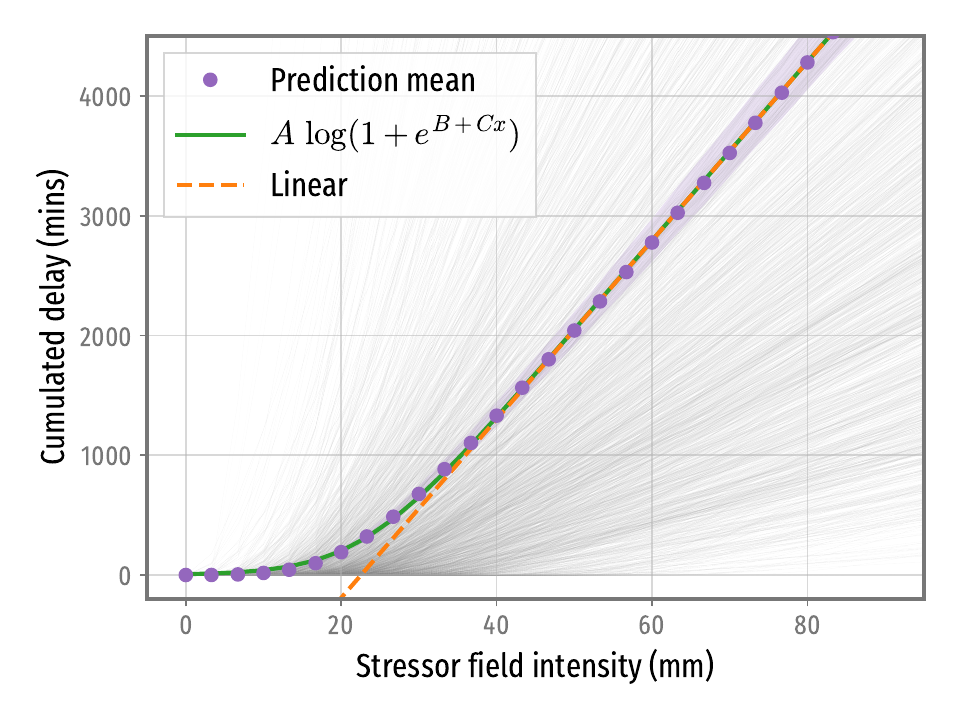} &
			\includegraphics[width=0.49\textwidth]{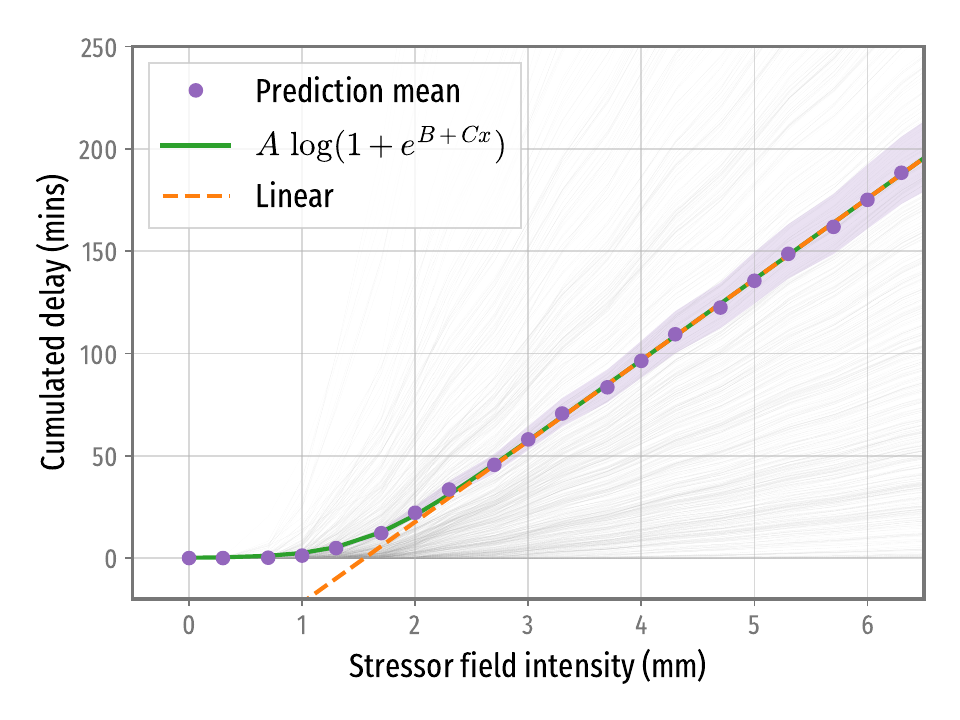}
		\end{tabular}
	\end{center}
	\caption{
		\textbf{Cumulated daily delays for a constant external stressor field.}
		The mean delay (dots) among all stations (light gray lines) are fitted to a log-exp function (green solid line).
		The linear limit at higher values of the external stressor field is reported as an orange dashed line.
	}\label{fig:daily_delay}
\end{figure}

Figure~\ref{fig:daily_delay} shows the response of the coupled system to a constant external stressor field at different intensity values.
The cumulative daily delays follow the following curve:
\begin{equation*}
	f(x) = A\log(1+e^{B+Cx})
\end{equation*}
that continuously combine two phases: an initial phase at small values of the external stressor field where the delay is easily absorbed by the system; and one at higher intensities of the external stressor field where the delay linearly increases.

\newpage
\subsection{Ridge Regression}

Let us predict the excess of delay based on the ridge regression of the datasets.

\begin{figure}[!htb]
	\begin{center}
		\includegraphics[width=0.45\textwidth]{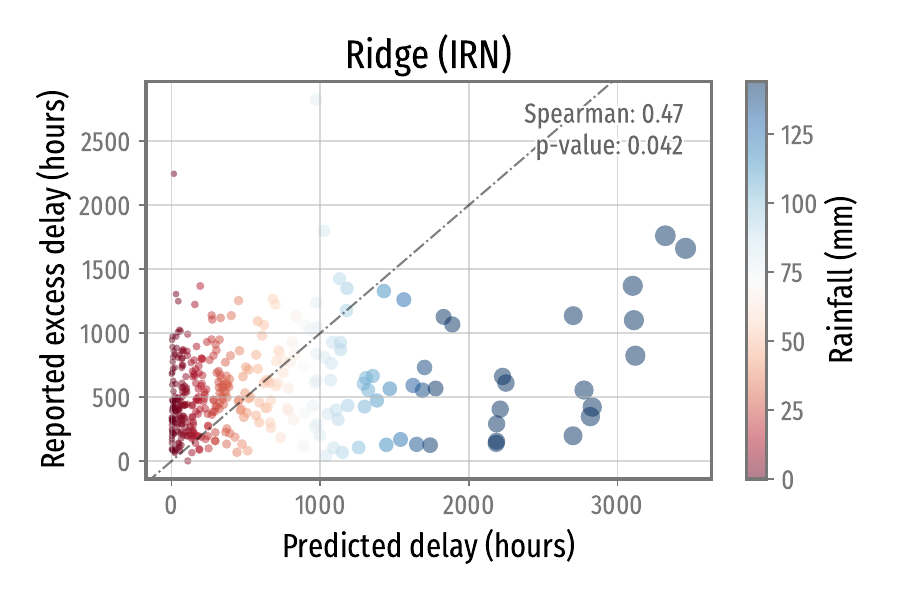}
		\includegraphics[width=0.45\textwidth]{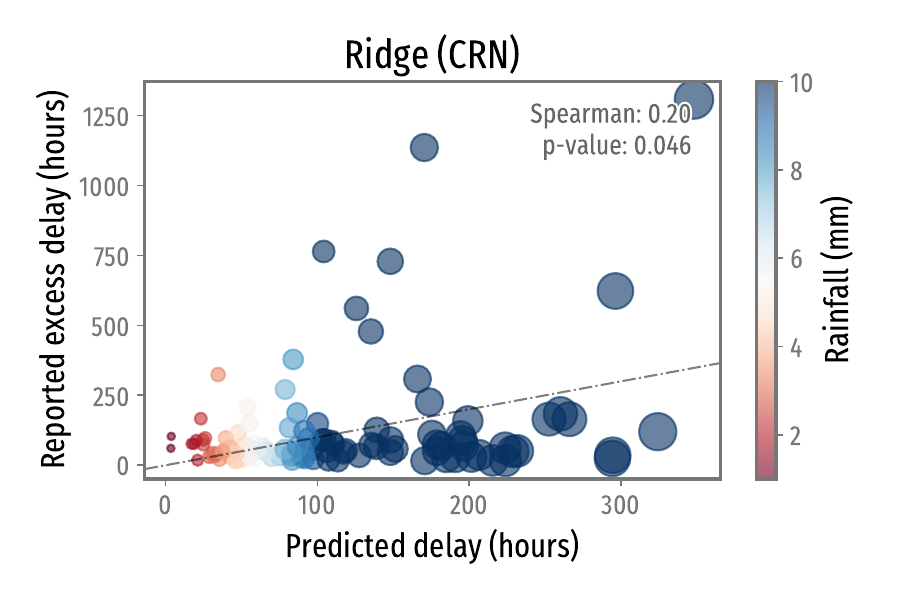}
	\end{center}
	\caption{
		\textbf{Baseline model for comparison.}
		Excess delay as predicted by a Ridge regressor on the Italian (left) and Chinese (right) datasets.
	}\label{fig:regression}
\end{figure}

Figure~\ref{fig:regression} displays the prediction of a Ridge regressor and the correlation of this with the reported excess of delay.

\newpage
\subsection{CNR steady state}

\begin{figure}[!htb]
	\begin{center}
		\includegraphics[width=0.95\textwidth]{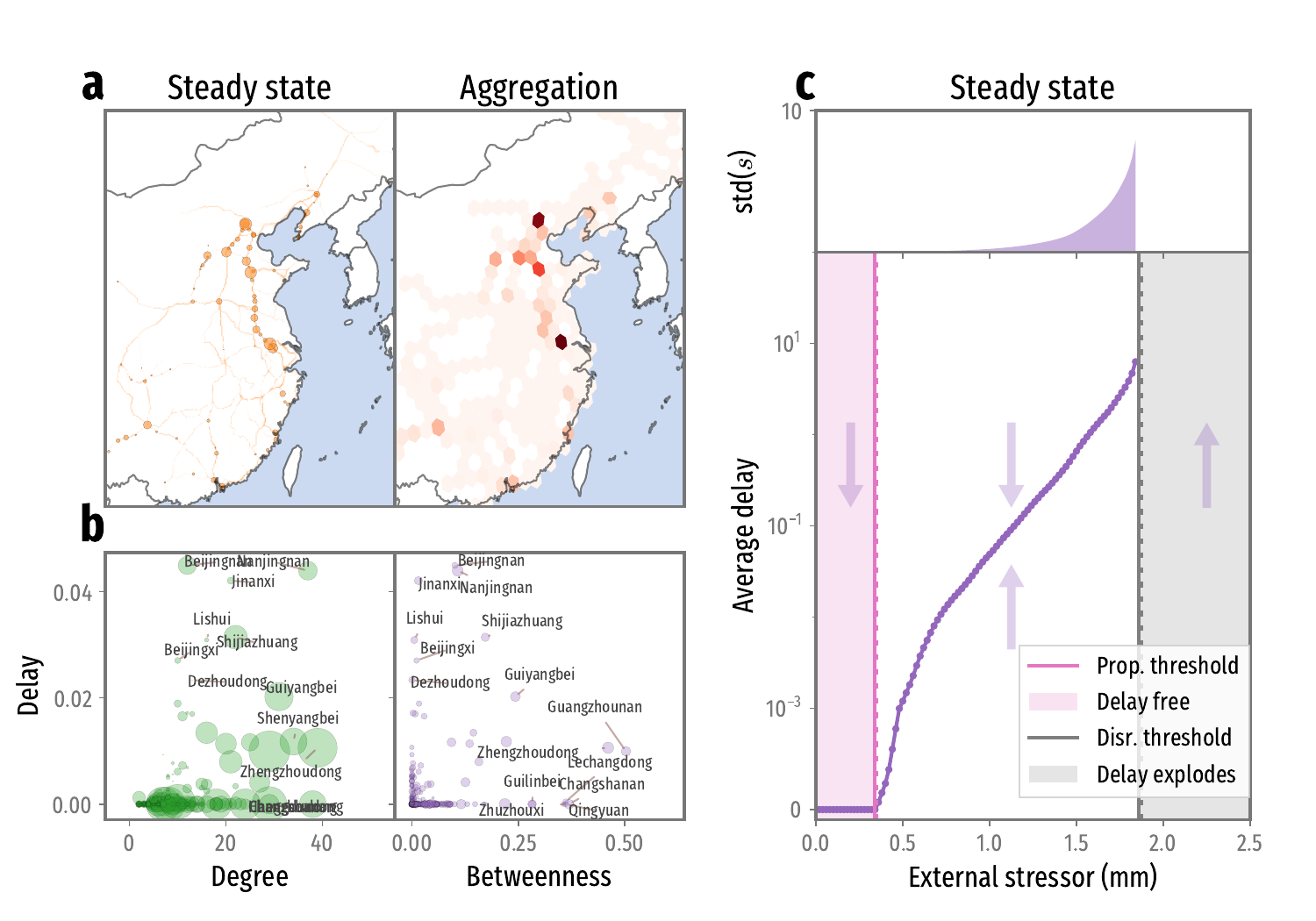}
	\end{center}
	\caption{
		\textbf{Steady state distribution for CRN}.
	}\label{fig:crn_steadystate}
\end{figure}

In this Figure we report the steady state analysis for the CRN dataset.
The results are qualitatively similar to those for the IRC although the thresholds are lower.
This is due to the lower levels of rainfall stress to which Chinese railways are exposed.

\newpage
\subsection{Empirical evidence of propagation threshold}

\begin{figure}[!htb]
	\begin{center}
		\includegraphics[width=0.95\textwidth]{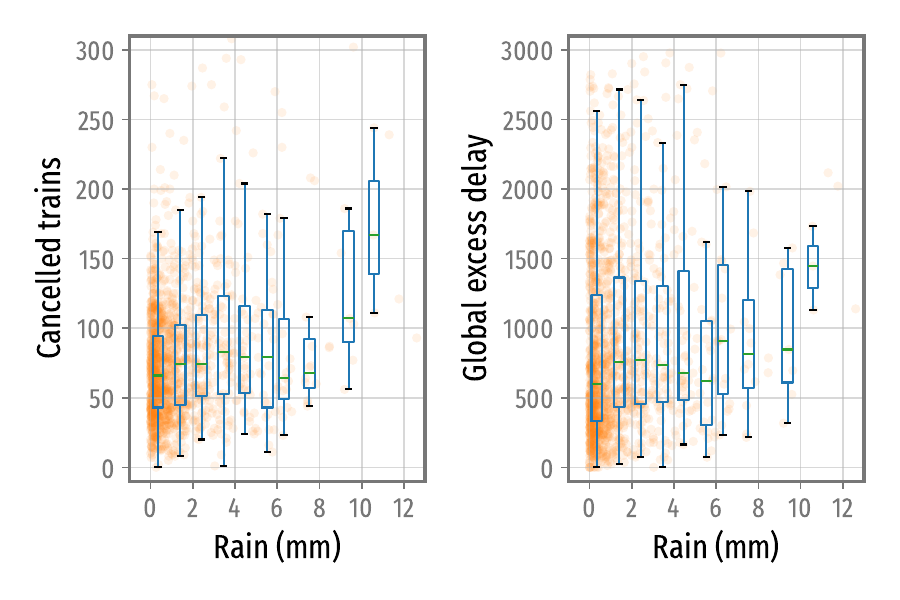}
	\end{center}
	\caption{
		\textbf{Number of trains cancelled and excess of delay versus rainfall}.
	}\label{fig:proxy}
\end{figure}

In Figure~\ref{fig:proxy}, one can notice that the number of trains canceled and the global excess of delay start to increase after a given value of the intensity of the rainfall.
This empirical observation may support the existence of the \emph{propagation threshold}.

\end{document}